\documentclass[onecolumn,12pt]{article}

\usepackage{jheppub}
\usepackage{ifpdf}
\usepackage[bb=boondox]{mathalfa}

\graphicspath{ {figures/} }

\usepackage{amsfonts}
\usepackage{latexsym}
\usepackage{color}
\usepackage{bm}

\usepackage{amsmath,braket}
\usepackage{amssymb,physics}
\usepackage{amsthm}
\usepackage{mathtools}

\usepackage{epsfig}

\usepackage{tensor}
\usepackage{pdfpages}
\usepackage{bbm}
\usepackage{graphicx,epstopdf}
\usepackage[numbers]{natbib} 
\usepackage[makeroom]{cancel}
\usepackage{hyperref}
\usepackage{array}
\usepackage[export]{adjustbox}

\usepackage[normalem]{ulem}
\usepackage{romannum}

\usepackage{ulem}

\usepackage{tikz}
\usetikzlibrary{intersections}
\usepackage{circuitikz}

\usepackage{subcaption}

\numberwithin{equation}{section}									% equation numbering by section

\usepackage{here}

\usepackage{comment}

\newcommand{\bea}{\begin{eqnarray}}
	\newcommand{\eea}{\end{eqnarray}}
\newcommand{\bes}{\begin{equation*}}
	\newcommand{\beas}{\begin{eqnarray*}}
		\newcommand{\eeas}{\end{eqnarray*}}
	\newcommand{\bas}{\begin{array*}}
		\newcommand{\eas}{\end{array*}}
	\newcommand{\ees}{\end{equation*}}

\subheader{\today}
\title{\boldmath c-theorem and improvement in non-compact conformal field theories}

\author[a]{Nanami Nakamura,}
\author[a]{Yu Nakayama}

\affiliation[a]{Center for Gravitational Physics and Quantum Information, Yukawa Institute for Theoretical Physics, Kyoto University,\\
	Kitashirakawa Oiwakecho, Sakyo-ku, Kyoto 606-8502, Japan}
\emailAdd{nanami.nakamura@yukawa.kyoto-u.ac.jp}
\emailAdd{yu.nakayama@yukawa.kyoto-u.ac.jp}

\abstract{Energy-momentum tensor in general conformal field theories have improvement ambiguity and it can affect the argument in deriving c-theorem. While the derivation of Zamolodchikov's c-theorem is still formally valid with the improved energy-momentum tensor, the behavior of the c-function proposed by Zamolodchikov becomes different. When the theory suffers an IR divergence due to the non-compactness, the c-function can be unbounded and may even show non-monotonicity. We find that one of the three-point function sum rule proposed by Hartman and Mathys motivated by the averaged null energy condition, however, is agnostic about the improvement and gives the (effective) Virasoro central charge, when the IR theory is gapped.}

\begin{document} 
	
%%%%%%%%%%%%%%%%%%%%%%%%%%%%%%%%%%%%%%%%%%%%%%%%%%%%
\begin{flushright}
YITP-25-37
\\
\end{flushright}
%%%%%%%%%%%%%%%%%%%%%%%%%%%%%%%%%%%%%%%%%%%%%%%%%%%%
\maketitle
\flushbottom

%%%%%%%%%%%%%%%%%%%%%%%%%%%%%%%%%%%%%%%%%%%%%%%%%%%%%%%%
%%%%%%%%%%%%%%%%%%%%%%%%%%%%%%%%%%%%%%%%%%%%%%%%%%%%%%%%
\section{Introduction}
Contrary to its name, a renormalization group (RG) is not a group, but it is a semi-group.\footnote{Or more pedantically it is a monoid because of the existence of the identity.} This physical intuition is beautifully demonstrated by the c-theorem that there exists a number, Virasoro central charge in two-dimensional conformal field theories (CFTs), which always decreases from the ultraviolet (UV) fixed point to the infrared (IR) fixed point if the two theories are connected by an RG flow.

In the original derivation by Zamolodchikov \cite{Zamolodchikov:1986gt}, the monotonically decreasing c-function was introduced. As later pointed out by Polchinski \cite{Polchinski:1987dy}, while the derivation of monotonicity is formally valid, the c-function suffers an ambiguity because the energy-momentum tensor, used in the definition of the c-function, has improvement ambiguity in general quantum field theories. Polchinski showed that there should be a recipe to pick up a particularly good energy-momentum tensor that gives a monotonic and finite c-function that coincides with the Virasoro central charges at the fixed point.

In practice, it is not immediately obvious how to implement this idea practically before we know everything about the RG flow. How can we tune the IR energy-momentum tensor within the UV definition when we do not know what happens in the IR limit? It sounds quite challenging. Furthermore, the other physical principles, such as shift symmetry in the Nambu-Goldstone bosons although this does not occur in two dimensions, may prevent us from improving the energy-momentum tensor as we wish. The situation becomes most complicated when the conformal field theory is non-compact, where the improvement program by Polchinski might not work. Such examples include \cite{Hull:1985rc,Itsios:2021eig}.

In this paper, we study the effect of the improvement of the energy-momentum tensor in the RG flow of non-compact quantum field theories in two dimensions, taking a free non-compact scalar as a canonical example. Depending on the choice of the improvement terms in the energy-momentum tensor, we observe that the c-function proposed by Zamolodchikov behaves very differently; sometimes it just changes the fixed point central charges, sometimes it is monotonic but unbounded, or sometimes even the monotonicity is violated due to the IR divergence.

This begs us the question of whether there is any better quantity than Zamolochikov's c-function. We find that one of the three-point function sum rule proposed by Hartman and Mathys \cite{hartman2023twodimensional} may be one such
quantity. This sum rule is motivated by the positivity of the averaged null energy condition \cite{Faulkner:2016mzt,Hartman:2016lgu}, and the relation to the c-function was proved there (with assumptions that are not necessarily valid in our setup). We show that it is agnostic about the improvement and gives the (effective) Virasoro central charge when the IR theory is gapped.

The organization of the paper is as follows. In section \ref{sec:zamo}, we study the effect of the improvement term in the original discussions of c-theorem by Zamolodchikov. In section \ref{sec:2ptexample}, we study the explicit example of a free scalar with or without mass term to see how Zamolodchikov's c-function will behave differently due to the improvement term. In section \ref{sec:3ptsum}, we study the effect of the improvement term in the three-point function sum rule proposed by Hartman and Mathys. In section \ref{sec:3ptexmple}, we demonstrate the three-point function sum rule in a free scalar theory with the improvement term. In section \ref{sec:discussion}, we conclude with some discussions.

\section{Zamolodchikov's c-theorem and improvement}\label{sec:zamo}
Our starting point is Zamolodchikov's original derivation of the c-theorem in general field theories in two dimensions. With a given action $S_E$ in curved space (with Euclidean signature), we may define the energy-momentum tensor by 
\begin{align}
T_{\mu\nu} &= -\frac{4\pi}{\sqrt{g}}\frac{\delta S_E}{\delta g^{\mu\nu}} . \label{EMT}
\end{align}
Introducing the trace $\Theta = T^{\mu}_{\mu} = 4 T_{z \bar{z}}$, we define the two-point functions $F$, $G$ and $H$ as \cite{Polchinski:1998rr}
\begin{align}
\label{FGH}
  \ev{T_{zz}(z,\bar{z}) T_{zz}(0)} &= \frac{1}{z^4} F(z\bar{z}M^2) ,\\
  \ev{\Theta(z,\bar{z}) T_{zz}(0)} &= \frac{1}{z^3 \bar{z}} G(z\bar{z}M^2) ,\\
  \ev{\Theta(z,\bar{z}) \Theta(0)} &= \frac{1}{z^2 \bar{z}^2} H(z\bar{z}M^2). 
\end{align}
Here $M >0$ is a reference energy scale.

In a classic paper \cite{Zamolodchikov:1986gt}, Zamolodchikov showed that a particular linear combination
\begin{align}
\label{c_fn}
  C = 2F - G - \frac{3}{8}H
\end{align}
is monotonically decreasing along the RG flow. This is because the conservation of the energy-momentum tensor gives relations between scale derivative $\dot{C} = \frac{\partial}{\partial \log (z\bar{z})} C$ of two-point functions
\begin{align}
\label{con_aa}
  4\dot{F} + \dot{G} -3G &= 0 \cr
  4\dot{G} - 4G + \dot{H} - 2H &= 0  
\end{align}
and
\begin{align}
\label{con_bb}
    \dot{C} = -\frac{3}{4}H,
\end{align}
so the monotonicity $\dot{C} \le 0$ follows from the unitarity. 

In a more recent paper \cite{hartman2023twodimensional}, Hartman and Mathys showed that $F$ itself is monotonic from the positive spectrum axiom. As they emphasized, the positive spectrum axiom is closely related to the averaged null energy condition in two dimensions.

So far, we have not assumed anything about the energy-momentum tensor except that it is symmetric and conserved. In particular, it applies to arbitrary improved energy-momentum tensor. Suppose we have an arbitrary scalar operator $O$, we can redefine the energy-momentum tensor 
\begin{align}
T_{\mu\nu} \to T_{\mu\nu} + \beta (\partial_\mu \partial_\nu -g_{\mu\nu} \partial_\mu \partial^\mu) O \ . 
\end{align}
which is still symmetric and conserved. A more complicated improvement with a symmetric tensor is possible but it should be inconsistent with the unitarity (see e.g. \cite{Polchinski:1987dy} or a review \cite{Nakayama:2013is}). 
Apriori, we do not know which energy-momentum tensor is used in a given RG scheme.

Of course, the best choice if possible is to select the RG scheme so that the energy-momentum tensor at the UV and IR fixed point is both traceless. This choice is particularly good because the $C$ becomes constant and can be identified with the Virasoro central charges. Moreover, we can show the sum rule \cite{Cappelli:1990yc} %\comments{Normalization not fixed yet.}
\begin{align}
c_{\text{UV}} - c_{\text{IR}} = \frac{3\pi}{(2\pi)^2}\int d^2x x^2 \langle \Theta(x) \Theta(0) \rangle_{\text{sep}} \label{Zsum}
\end{align}
if $\Theta_{\text{IR}} = \Theta_{\text{UV}} = 0$ as an operator identity.\footnote{This assumption implies that the fixed points are conformal invariant rather than just scale invariant. Polchinski \cite{Polchinski:1987dy} argued that there should be an RG scheme that \eqref{Zsum} must be finite if the theory is compact (there was a small sign error in the original paper: see e.g. \cite{Dymarsky:2013pqa,Nakayama:2013is} for some details).  Then unitarity should demand $\Theta = 0$ at the fixed points.} The operator identity $\Theta = 0$ still allows contact terms in correlation functions of $\Theta$, and $|_{\text{sep}}$ means we discard the contact term. This sum rule can be derived in various ways. The one based on the dilaton Wess-Zumino action can be found in \cite{Komargodski:2011xv}.
See also \cite{Loparco:2024ibp,Abate:2024xyb} for a generalization in de-Sitter space.

As mentioned in the introduction, the other symmetry may not allow an arbitrary choice, or more practically, implementing this particular choice can be difficult unless we have solved everything as in the integrable flow. This is because the choice requires a shooting method specified by the UV limit and the IR limit.

On the other hand, the argument by Zamolodchikov should be still valid with any improvement term at least formally. There are a couple of issues here. One issue is that the right-hand side of the sum rule \eqref{Zsum} may become divergent and the interpretation becomes less obvious. A further issue is that if the theory under consideration is non-compact, the severe IR divergence can cause the violation of the positivity. We will inspect these issues with a simple non-compact free scalar theory (with or without mass term) in the next section.

\section{A free scalar with arbitrary improvement terms: two-point functions}\label{sec:2ptexample}

We examine the effect of the improvement term on the c-theorem in the simplest non-compact field theory in two dimensions i.e. a free real scalar. The action in the Euclidean signature is given by
\begin{align}
    S= \frac{1}{2\pi\alpha'}\int dzd\bar{z} \qty[\partial X\bar{\partial} X+\frac{1}{2}m^2X^2]
\end{align}

Our convention follows the one used in the string theory textbook by Polchinski \cite{Polchinski:1998rq,Polchinski:1998rr}.
This theory is free so we can compute all the correlation functions by the Wick theorem. Our propagator $\mathcal{G}(z)$ is normalized to be 

\begin{align}
\mathcal{G} = \alpha' K_0(m|z|) \ ,
\end{align}
so that $(\partial \bar{\partial} +\frac{1}{4}m^2) \mathcal{G} = -\pi \alpha' \delta(z, \bar{z})$.\footnote{We recall that our convention is $dz d\bar{z} = 2 d^2x = 2dx_1 dx_2$ and $\delta(z,\bar{z}) = \frac{1}{2} \delta(x_1) \delta(x_2)$, where we use $z=x_1+ix_2, \bar{z}=x_1-ix_2$.} %\comments{Make the equation correct...}

In the massless limit, we use
\footnote{The massless scalar propagator in two-dimension is famously negative in the large $|z|$ due to the IR divergence. In this sense, the unitarity or reflection positivity is subtly violated due to the IR regularization. Often in the literature, this violation is neglected by focusing on the vertex operator which has positive two-point functions, or by compactifying the scalar with a large radius. In this paper, we do study this subtle issue more seriously.}
\begin{align}
\mathcal{G} = -\alpha' \log(M |z|).
\end{align}
To match the UV behavior, we recall the Taylor expansion of the Bessel function:
\begin{align}
  K_0(x) = -\gamma + \log2 - \log x + \frac{1}{4}(1 -\gamma + \log2 - \log x)x^2 + \cdots,
\end{align}
where  $\gamma = 0.577216\cdots$ is the Euler gamma, so we  will set $M = \frac{e^{\gamma}}{2} m$.

To demonstrate the c-theorem, we need the energy-momentum tensor, which is defined in \eqref{EMT}. In the free scalar theory, we can add a curvature coupling (or dilaton coupling in string theory) $\beta R f(X)$, and the energy-momentum tensor has a degree of freedom in the improvement term: 
\begin{align*}
  T^{(\beta)}_{\mu\nu} &= -\frac{1}{\alpha'}\qty[\partial_\mu X \partial_\nu X - g_{\mu\nu}\qty(\frac{1}{2}\partial_\rho X \partial^\rho X + \frac{1}{2}m^2X^2)
  + \beta (\partial_\mu \partial_\nu - g_{\mu\nu }\partial^\mu \partial_\mu) f(X)] \\
  {T^{(\beta)}}_\mu^\mu &= \frac{1}{\alpha'} \qty[m^2 X^2 + \beta \partial^\rho\partial_\rho f(X)].
\end{align*}
Clearly the improvement term proportional to $\beta$ is conserved independently.\footnote{This ambiguity of the energy-momentum tensor that follows from the non-minimal coupling to the gravity, is historically known as improvement \cite{Callan:1970ze,Polchinski:1987dy} because we can use the freedom to redefine the energy-momentum tensor with desired properties such as finineness, tracelessness or shift-symmetry.}

Swiching to the complex coordinate $g_{zz} = 0, g_{z\bar{z}} = \frac{1}{2}$ with $\partial = \partial_z, \bar{\partial} = \partial_{\bar{z}}$, we have
\begin{align}
  T_{zz}^{(\beta)} &= -\frac{1}{\alpha'}\qty[\partial X \partial X + \beta \partial \partial f(X)] \\
  \Theta^{(\beta)} &= 4T_{z\bar{z}}^{(\beta)} = \frac{1}{\alpha'}[ m^2 X^2 + 4 \beta \partial \bar{\partial} f(X)]. 
\end{align}
In these expressions, normal ordering is implicitly assumed.

We now study the two-point functions of the energy-momentum tensor with particular choices of $f(X) = X^2 $ and $X^3$. 

When $f(X)=X^2$, we have
\begin{align}
\label{2pt_X^2}
  \ev{T_{zz}^{(\beta)}(z_1) T_{zz}^{(\beta)}(z_2)}
  =&\frac{1}{\alpha'^2}\qty[2(\partial_1\partial_2 \mathcal{G})^2 + 8\beta\qty[(\partial_1\partial_2 \mathcal{G})^2
  + (\partial_1 \mathcal{G})(\partial_1\partial_2\partial_2 \mathcal{G})] +2 \beta^2\partial_1\partial_1\partial_2\partial_2 \mathcal{G}^2] \nonumber\\
  \ev{T_{zz}^{(\beta)}(z_1) \Theta^{(\beta)}(z_2)}
  =&\frac{1}{\alpha'^2}[-2m^2(\partial_1 \mathcal{G})^2 \nonumber\\
  &\quad- \beta\qty(16(\partial_1\partial_2\mathcal{G})(\partial_1\bar{\partial}_2\mathcal{G})+2m^2\partial_1\partial_1\mathcal{G}^2+4m^2(\partial_1\mathcal{G})^2)-8\beta^2\beta^2\partial_1\partial_1\partial_2\bar{\partial}_2 \mathcal{G}^2]\nonumber\\
  \ev{\Theta^{(\beta)}(z_1) \Theta^{(\beta)}(z_2)}
  =&\frac{1}{\alpha'^2}\qty[2m^4 \mathcal{G}^2 + 16\beta m^2\partial_1\bar{\partial}_1\mathcal{G}^2+32\beta^2 \partial_1\bar{\partial}_1 \partial_2\bar{\partial}_2 \mathcal{G}^2].
\end{align}
When $f(X)=X^3$, we have
\begin{align}
\label{2pt_X^3}
  \ev{T_{zz}^{(\beta)}(z_1) T_{zz}^{(\beta)}(z_2)}
  =&\frac{1}{\alpha'^2}\qty[2(\partial_1\partial_2 \mathcal{G})^2
  +6 \beta^2\partial_1\partial_1\partial_2\partial_2 \mathcal{G}^3] \nonumber\\
  \ev{T_{zz}^{(\beta)}(z_1) \Theta^{(\beta)}(z_2)}
  =&\frac{1}{\alpha'^2}\qty[-2m^2(\partial_1 \mathcal{G})^2 -24\beta^2\partial_1\partial_1\partial_2\bar{\partial}_2\mathcal{G}^3]\nonumber\\
  \ev{\Theta^{(\beta)}(z_1) \Theta^{(\beta)}(z_2)}
  =&\frac{1}{\alpha'^2}\qty[2m^4 \mathcal{G}^2 
  + 6\cdot16\beta^2 \partial_1\bar{\partial}_1 \partial_2\bar{\partial}_2 \mathcal{G}^3].
\end{align}
Note that there is no order $\beta$ term here because correlation functions with odd numbers of $X$ vanish.

%%%%%%%%%%%%%%%%% massless
\subsection{massless case}\label{subsec:massless}
When $m^2 = 0$, the theory must be scale invariant, but it may suffer IR divergence. 
This is because the propagator $\mathcal{G}(z,0) = -\alpha' \log (M |z|)$ becomes negative for large separation $|z|$ while the unitarity should demand the two-point function must be positive. 
Accordingly, the c-function defined by Zamolodchikov may depend on the scale in a hidden manner. Let us see this more explicitly here.

%%%%%%%%%%%%%%%%%% 二乗
\subsubsection{\underline{Example 1: $f(X) = X^2$}}
In this case, evaluating \eqref{2pt_X^2} gives
\begin{align}
  F &= \frac{1}{2} + 6\beta + (11 - 12\log x)\beta^2 \nonumber\\
  G &= -8 \beta^2 \nonumber\\
  H &= 16 \beta^2, 
\end{align}  
and from (\ref{c_fn}) Zamolodchikov's c-function becomes
\begin{align}
  C &= 1 + 12\beta + (24 - 12\log x)\beta^2,
\end{align}
where we put $x=M|z_{12}|$. They satisfy the consistency conditions (\ref{con_aa}) and (\ref{con_bb}), but they are non-trivial functions of $x$. 

%%%%%%%%%%%%%%%%%%%%%%
Unlike the case without the improvement term (i.e. $\beta=0$, Figure \ref{fig:No_improvement}),  $F$ and $C$ are not constant even though the theory is supposed to be scale (or conformal) invariant and there should be no RG flow. In particular, note that while $C$ is monotonically decreasing, it becomes negative at larger $|z|$. See the left panel of Figure \ref{fig:massless}.
Again, this is due to the IR divergence of the non-compact scalar. Also, if we tried to read the Virasoro central charge from the UV behavior of $C$ (or $F$), we would conclude it becomes infinite while the Virasoro central charge of a free real scalar without improvement is just $1$. We cannot read the Virasoro central charge from $C$ or $F$ when the energy-momentum tensor is not traceless.

%%%%%%%%%%%%%%%%%% 三乗
\subsubsection{\underline{Example 2: $f(X) = X^3$}}

\begin{figure}[h]
  \centering
  \begin{minipage}{0.49\columnwidth}
  \includegraphics[width=\columnwidth]{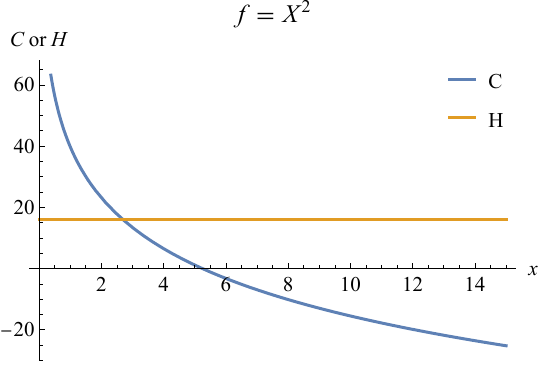}
  \end{minipage}
  \begin{minipage}{0.49\columnwidth}
     \centering
     \includegraphics[width=\columnwidth]{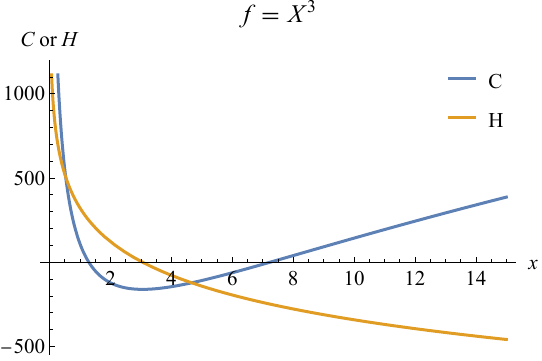}
  \end{minipage}
  \caption{The plots of $C(x)$ and $H(x)$ when $f(X)=X^2$ (left) and $f(X)=X^3$ (right). We set $\alpha'=2$ and $\beta = 1$. In the right plot ($f(X)=X^3$), $C(x)$ ceases to be monotonic when $H(x)<0$.}
  \label{fig:massless}
\end{figure}

This case is more interesting. 
The two-point functions (\ref{2pt_X^3}) become
\begin{align}
  F &= \frac{1}{2} - \frac{9}{2} \alpha' \beta^2 (-12 (\log x)^2 + 22 \log x - 3)\nonumber\\
  G &= 18 \alpha' \beta^2 (4 \log x - 3) \nonumber\\
  H &= - 144 \alpha' \beta^2 (\log x - 1)
\end{align}
and Zamolodchikov's c-function (\ref{c_fn}) now reads
\begin{align}
  C &= 1 - 27 \alpha' \beta^2 (-4(\log x)^2 + 8\log x - 1).
\end{align}
They satisfy the consistency conditions (\ref{con_aa}) and (\ref{con_bb}). We realize that the c-function is no longer a monotonic function! See the right panel of Figure \ref{fig:massless}. This is because $H$ is not positive for larger $|z|$ (Compare the blue and orange lines in the right panel of Figure \ref{fig:massless}). This comes from the IR divergence of the propagator and the energy-momentum tensor shows the subtle violation of the reflection positivity (or unitarity) mentioned in footnote 4.
We clearly see that the IR divergence and the improvement of the energy-momentum tensor may alter the fundamental property of Zamolodchikov's c-function.

%%%%%%%%%%%%%%%%%%%%%% massive
\subsection{massive case}\label{subsec:massive}
So far we have studied the effect of the improvement term in Zamolodchikov's c-function at the fixed point of a free massless scalar. Let us now study the simple RG flow by introducing a non-zero mass term. We expect that the introduction of mass will remove the IR divergence of the non-compact scalar and the pathological IR behavior of Zamolodchikov's c-function we saw in section \ref{subsec:massless} will disappear.

%%%%%%%%%%%%%%%%%%　二乗
\subsubsection{\underline{Without the improvement term}}
\begin{figure}[h]
  \centering
  \begin{minipage}{0.49\columnwidth}
  \includegraphics[width=\columnwidth]{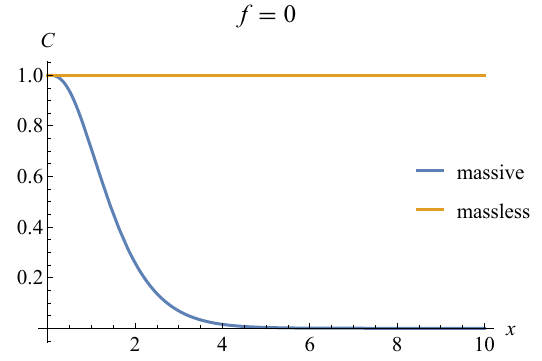}
  \end{minipage}
  \begin{minipage}{0.49\columnwidth}
     \centering
     \includegraphics[width=\columnwidth]{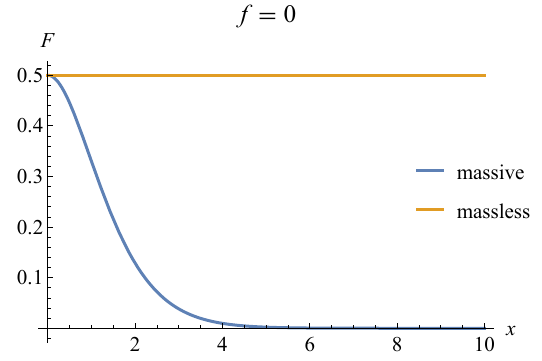}
  \end{minipage}
  \caption{The plots of $C(x)$ (left) and $F(x)$ (right) when $f(X)=0$. We set $\alpha'=2$.}
  \label{fig:No_improvement}
\end{figure}

For reference, we quote the two-point functions of the energy-momentum tensor without improvement term: 
\begin{align}
  F = \frac{1}{8}x^4 K_2(x)^2 ,\quad
  G = -\frac{1}{2}x^4 K_1(x)^2 ,\quad
  H = 2 x^4 K_0(x)^2,
\end{align}
where we put $x=m|z_{12}|$. Zamolodchikov's c-function (\ref{c_fn}) becomes
\begin{align}
    C = \frac{1}{4} x^4 \left(-3 K_0(x){}^2+2 K_1(x){}^2+K_2(x){}^2\right).
\end{align}
They satisfy (\ref{con_aa}) and (\ref{con_bb}). We show the comparison of $C(x)$ and $F(x)$ between the massive case and massless case in Figure \ref{fig:No_improvement}.

\subsubsection{\underline{Example 1: $f(X)=X^2$}}
With the use of the Bessel function identities, the two-point functions of the energy-momentum tensor become 
\begin{equation}
\begin{split}
  F&=\frac{x^4}{8} K_2(x){}^2 \\
  &+\frac{\beta  x^2} {8} (3 x^2 K_0(x){}^2 +(4 (x^2+11) K_1(x){}^2+x^2  K_2(x){}^2)  +28 x K_1(x) K_0(x))\\
  &+\frac{\beta^2 x}{16}(x (13 x^2+96) K_0(x){}^2 +12 (11 x^2+16)  K_1(x)  K_0(x) \\
  &\quad\quad\quad\quad\quad\quad+x (4 (4 x^2+41) K_1(x){}^2 +3 x^2 K_2(x){}^2))\\
  G&=-\frac{x^4 K_1(x){}^2}{2}\\
  &-\frac{\beta x^2}{4}(7 x^2 K_0(x){}^2+(8 x^2-4) K_1(x){}^2 +x^2 K_2(x){}^2+12 x K_1(x) K_0(x))\\
  &-\frac{\beta^2 x^2}{4} (13 x^2 K_0(x){}^2+4 (4 x^2+5)  K_1(x){}^2 +3 x^2 K_2(x){}^2 +36 x K_1(x) K_0(x))\\
  H&=2 x^4 K_0(x)^2\\
  &+ 8\beta x^4 (K_0(x){}^2 +K_1(x){}^2)\\
  &+ \beta^2 x^2(13 x^2 K_0(x){}^2 +{4 (4 x^2+1) K_1(x)^2} +3 x^2 K_2(x)^2+{4x K_1(x) K_0(x)})   .
\end{split}
\end{equation}
Zamolodchikov's c-function (\ref{c_fn}) can be written;
\begin{equation}
\begin{split}
    C &= \frac{1}{4} x^4 \left(-3 K_0(x){}^2+2 K_1(x){}^2+K_2(x){}^2\right)\\
    &+\frac{1}{2} \beta  x^2 \left(-x^2 K_0(x){}^2+x^2 K_2(x){}^2+20 x K_1(x) K_0(x)+20 K_1(x)^2\right)\\
    &+12\beta^2 x \left(2 \left(x^2+1\right) K_1(x) K_0(x)+x K_0(x){}^2+2 x K_1(x){}^2\right).
\end{split}
\end{equation}
They satisfy the consistency condition (\ref{con_aa}) and (\ref{con_bb}). 

\begin{figure}[h]
  \centering
  \begin{minipage}{0.48\columnwidth}
     \centering
     \includegraphics[width=\columnwidth]{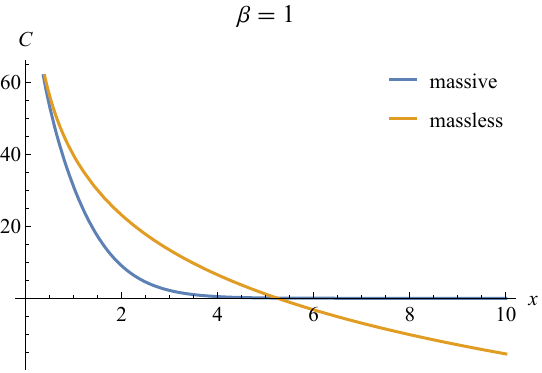}
  \end{minipage}
  \begin{minipage}{0.50\columnwidth}
    \centering
    \includegraphics[width=\columnwidth]{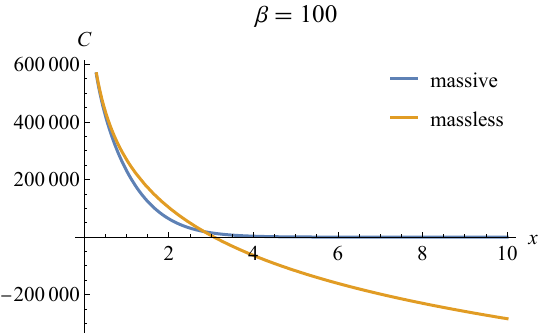}
  \end{minipage}
  \caption{The plots of $C(x)$ at $\beta=1$ (left), $\beta=100$ (right) when $f(X)=X^2$; the orange line is massless while the blue line is massive.} 
     \label{fig:cx2}
\end{figure}

\begin{figure}[h]
  \centering
  \begin{minipage}{0.48\columnwidth}
     \centering
     \includegraphics[width=\columnwidth]{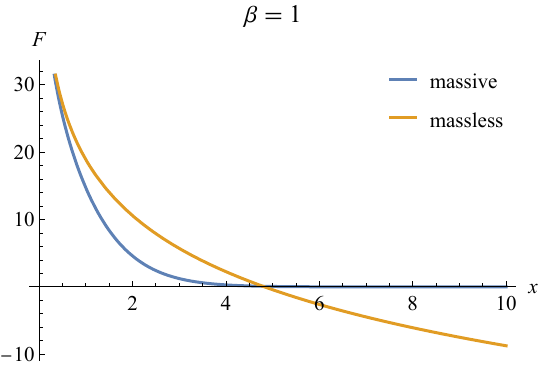}
  \end{minipage}
  \begin{minipage}{0.50\columnwidth}
    \centering
    \includegraphics[width=\columnwidth]{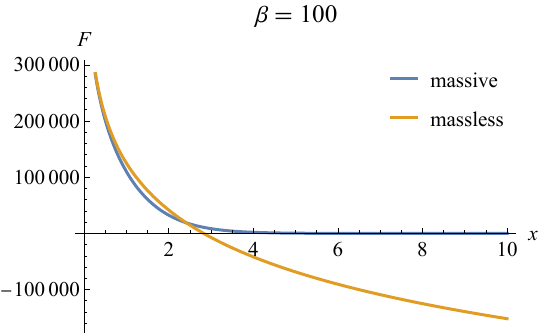}
  \end{minipage}
  \caption{The plots of $F(x)$ at $\beta=1$ (left), $\beta=100$ (right)  when $f(X)=X^2$; the orange line is massless while the blue line is massive.} 
     \label{fig:Fx2}
\end{figure}

We now compute Zamolodchikov's c-function and show the comparison of $C$ and $F$ between the massive case and massless case in Figure \ref{fig:cx2} and Figure \ref{fig:Fx2}. The behavior of $C$ and $F$ in the massless case and massive case agree in the UV limit. $C$ and $F$ are both monotonically decreasing as shown by Zamolodchikov and Hartman-Mathys. The IR behavior, however, is drastically different. In the massive case, $C$  and $F$ are always positive while the positivity was violated in the massless case. 
This is because the IR divergence is removed with the mass term and the positivity of the two-point function is guaranteed. 

Let us however stress that the sum rule \eqref{Zsum} is not valid with non-zero $\beta$. If we tried to integrate $H$ in the massive RG flow, we would get infinity while the central charge difference should be $1$.
This is because the assumption $\Theta_{\text{UV}} = 0$ is violated.

\subsubsection{\underline{Example 2: $f(X)=X^3$}}
Evaluating (\ref{2pt_X^3}) explicitly, we obtain
\begin{equation}
\begin{split}
    F(x) =& \frac{1}{8} x^4 K_2(x){}^2 \\
  &+\frac{9 \alpha' \beta^2 x}{8}  \left(x\left(7 x^2 +24\right)  K_0(x){}^3 +16 \left(4 x^2 +3\right) K_1(x) K_0(x){}^2\right.\\ &\left.\quad\quad\quad\quad\quad\quad\quad\quad+4x \left(5 x^2 +22\right) K_1(x){}^2  K_0(x) +24 x^2 K_1(x){}^3\right),\\
  G(x) =& -\frac{1}{2} x^4 K_1(x){}^2 \\
  &-\frac{9 \alpha' \beta^2 x^2}{2} \left(7 x^2 K_0(x){}^3 +4 \left(5 x^2+4\right) K_1(x){}^2 K_0(x) \right.\\ &\left.\quad\quad\quad\quad\quad\quad\quad\quad\quad\quad\quad +22 x K_1(x) K_0(x){}^2+12 x K_1(x){}^3\right),\\
  H(x) =&2 x^4 K_0(x)^2\\
  &+18 \alpha' \beta^2 x^2 \left(7 x^2 K_0(x){}^3 + 4 \left(5 x^2+2\right) K_1(x){}^2  K_0(x) \right.\\ &\left.\quad\quad\quad\quad\quad\quad\quad\quad\quad\quad\quad+8 x K_1(x) K_0(x){}^2 + 8 x K_1(x){}^3\right),
\end{split}
\end{equation}
which enables us to compute Zamolodchikov's c-function (\ref{c_fn}):
\begin{equation}
\begin{split}
  C(x) =& \frac{1}{4} x^4 \left(-3 K_0(x){}^2+2 K_1(x){}^2+K_2(x){}^2\right) \\
  &+27 \alpha' \beta ^2 x \left(\left(7 x^2+4\right) K_1(x) K_0(x){}^2+2 x^2 K_1(x){}^3 \right.\\
  &\left.\quad\quad\quad\quad\quad\quad+2 x K_0(x){}^3+8 x K_1(x){}^2 K_0(x)\right).
\end{split}    
\end{equation}  
We have checked the consistency conditions (\ref{con_aa}) and (\ref{con_bb}).

\begin{figure}[h]
  \centering
  \begin{minipage}{0.48\columnwidth}
     \centering
     \includegraphics[width=\columnwidth]{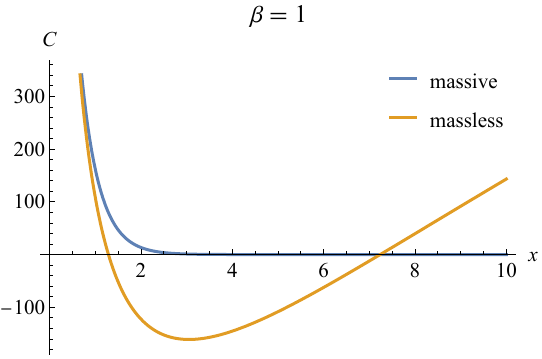}
  \end{minipage}
  \begin{minipage}{0.5\columnwidth}
    \centering
    \includegraphics[width=\columnwidth]{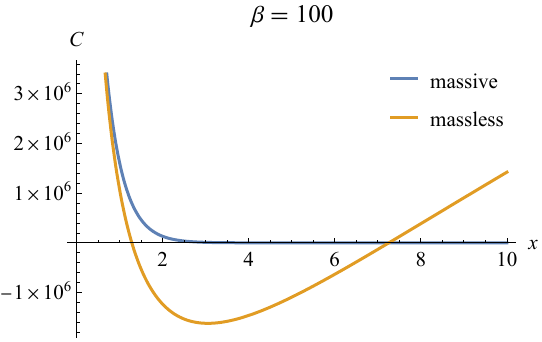}
  \end{minipage}
  \caption{The plots of $C(x)$ at $\beta=1$ (left), $\beta=100$ (right) when $f(X)=X^3$; the orange line is massless while the blue line is massive.} 
     \label{fig:cx3}
\end{figure}

\begin{figure}
    \centering
    \begin{minipage}{0.48\columnwidth}
     \centering
     \includegraphics[width=\columnwidth]{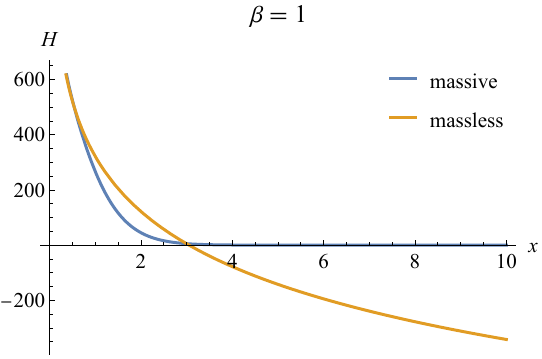}
  \end{minipage}
  \begin{minipage}{0.5\columnwidth}
    \centering
    \includegraphics[width=\columnwidth]{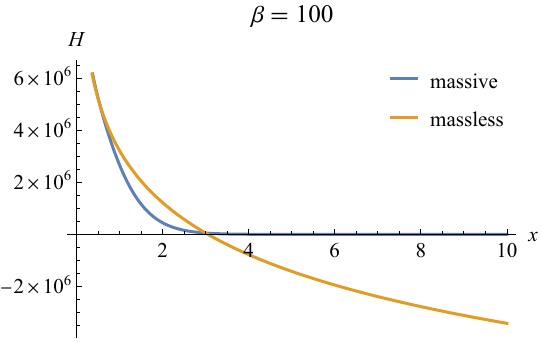}
  \end{minipage}
    \caption{The plots of $H(x)$ at $\beta=1$ (left) and $\beta=100$ (right) when $f(X) = X^3$; the orange line is massless while the blue line is massive.}
    \label{fig:Hx3}
\end{figure}

\begin{figure}[h]
  \centering
  \begin{minipage}{0.48\columnwidth}
     \centering
     \includegraphics[width=\columnwidth]{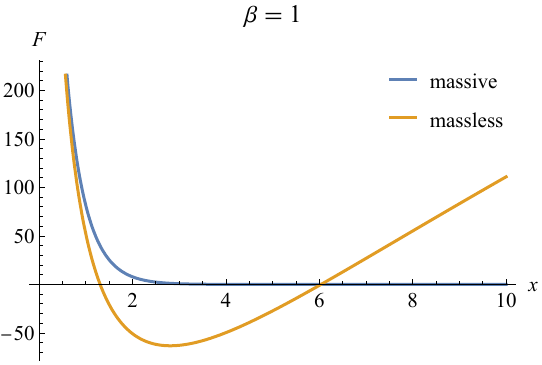}
  \end{minipage}
  \begin{minipage}{0.5\columnwidth}
    \centering
    \includegraphics[width=\columnwidth]{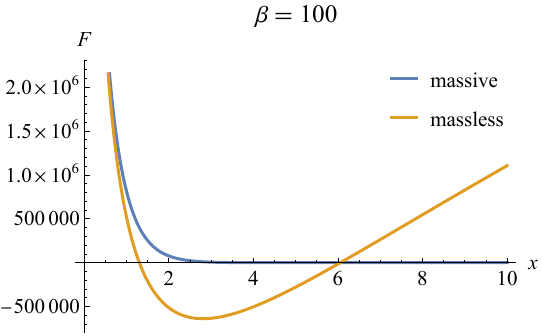}
  \end{minipage}
  \caption{The plots of $F(x)$ at $\beta=1$ (left), $\beta=100$ (right); the orange line is massless while the blue line is massive.} 
     \label{fig:Fx3}
\end{figure}

We plot $C$ in Figure \ref{fig:cx3}. In the massless case, we observed that $C$ is not monotonically decreasing, but in the massive flow, $C$ is monotonically decreasing as expected from Zamolodchikov's argument. The reason why $C$ was not monotonically decreasing in the massless case is the IR divergence causing non-positive $H$ as we can see from Figure \ref{fig:Hx3}. In the massive flow here, as we see in Figure \ref{fig:cx3} and Figure \ref{fig:Hx3}, due to the absence of the IR divergence, $H$ is always positive, and $C$ is monotonically decreasing. 

In Figure \ref{fig:Fx3}, we also plot $F$, which was proposed to be monotonically decreasing more recently in \cite{hartman2023twodimensional}. We again see that $F$ is not monotonically decreasing in the massless case due to the IR divergence, but becomes monotonically decreasing in the massive RG flow.

As in the $f(X) = X^2$ case, the massive RG flow does not change the UV behavior, so with the improvement, Zamolodchikov's sum rule \eqref{Zsum} is not valid. In the next section, we look for the other sum rule that may be valid even with the improved energy-momentum tensor.

\section{Hartman-Mathys three-point function sum rules}\label{sec:3ptsum}
In a beautiful paper \cite{hartman2023twodimensional}, motivated by the averaged null energy condition, Hartman and Mathys introduced the three-point function sum rules for the central charges. With the assumption of $\Theta_{\text{IR}} = \Theta_{\text{UV}} = 0$, they showed\footnote{Note that we follow Polchinski's notation \cite{Polchinski:1998rq}, so the definition of the energy-momentum tensor is different from the one used in \cite{hartman2023twodimensional} by a factor of $(-2\pi)$. In particular, $T_{uu}$ is classically {\it negative} in our notation.} 
\begin{align}
  c_{\mathrm{UV}} - c_{\mathrm{IR}}
  &= \frac{3\pi}{(2\pi)^3} \left( \partial_{k_{1u}} - \partial_{k_{2u}} \right)^2 
  \ev{\ev{\mathcal{T} [T_{uu}(-k_1 - k_2) \Theta(k_1) \Theta(k_2)]}}_{\text{sep}} \Big|_{k_1 = k_2 = 0} \ ,\label{Hsum}
\end{align}
where we use
\begin{align}
    \ev{\mathcal{O}(k_1)\cdots \mathcal{O}(k_n)} \coloneqq (2\pi)^d \delta^{(d)}(k_1+\cdots+k_n)\ev{\ev{\mathcal{O}(k_1)\cdots \mathcal{O}(k_n)}}.
\end{align}
From now on, we will work in the Minkowski signature and use the light-cone coordinate $u = t - x, v = t + x$. The metric is $ds^2 = -dudv$.

They have shown other various expressions for the same quantity such as
\begin{align}
\label{deltaC_X}
 c_{\mathrm{UV}} - c_{\mathrm{IR}} &= -\frac{3\pi}{(2\pi)^3} \int du_3 \int d^2x_1 d^2x_2 \, u_1^2 \delta(u_2) 
  \ev{\mathcal{T} [T_{uu}(u_3, v_3 = 0) \Theta(x_1)
  \Theta(x_2)]}_{\text{sep}} \cr
  &= \frac{6\pi}{(2\pi)^3} \int_{v_1<0,v_2<0} d^2x_1 d^2x_2 \, (u_1-u_2)^2 
  \ev{ \Theta(x_1) T_{uu}(0) 
  \Theta(x_2)}_{\text{sep}},
\end{align}
but when the assumption $\Theta_{\mathrm{IR}} = \Theta_{\mathrm{UV}} = 0$ is dropped to incorporate the effect of the improvement of the energy-momentum tensor, their equivalence may not hold due to the IR as well as UV divergence and the surface terms of the integration by part.

We, however, find the particular expression \eqref{Hsum} is interesting. We now show that it is independent of the improvement term as long as the IR theory is gapped (i.e. $c_{\mathrm{IR}} = 0$), and the UV divergence is mild enough.\footnote{One may be able to replace the condition that the IR theory is gapped by the condition that the IR theory is a {\it compact} CFT. The explicit demonstration of this by fine-tuning the potential $V(X)$ to reach a non-trivial Landau-Ginzburg fixed point is challenging though.}

Here we provide formal proof that it is independent of the improvement term. 
Consder the energy-momentum tensor with improvement term; $T^{(\beta)}_{\mu\nu} = T_{\mu\nu}^{(0)} + \beta(\partial_\mu \partial_\nu - g_{\mu\nu} \partial_\rho \partial^\rho)O$. 
Let us decompose the three-point functions we are interested in as
 \begin{align}
  &\ev{T_{uu}^{(\beta)}(x_3) \Theta^{(\beta)}(x_1) \Theta^{(\beta)}(x_2)} \nonumber\\
  &= \langle(T_{uu}^{(0)}(x_3) + \beta(\partial_u \partial_u)O(x_3))\nonumber\\
  &\quad\quad\quad\quad (\Theta^{(0)}(x_1) -4\beta\partial_{u1}\partial_{v1} O(x_1))(\Theta^{(0)}(x_2) -4\beta\partial_{u2}\partial_{v2} O(x_2))\rangle\nonumber\\
  &=\mathcal{A} 
  +\beta [-8 \partial_{u2} \partial_{v2} \mathcal{A}^* + \mathcal{B}] \nonumber\\
  &\quad\quad\quad\quad\quad
  + \beta^2[16 \partial_{u1} \partial_{v1} \partial_{u2} \partial_{v2} \mathcal{A}^{**}-8\partial_{u1} \partial_{v1} \mathcal{B}^*  ]
  + 16\beta^3 \partial_{u1} \partial_{v1} \partial_{u2} \partial_{v2} \mathcal{B}^{**},
\end{align}
where we have introduced 
\begin{align}
      \mathcal{A} &= \ev{T_{uu}^{(0)}(x_3) \Theta^{(0)}(x_1) \Theta^{(0)}(x_2)}, & \mathcal{B} &=\partial_{u3} \partial_{u3} \ev{O(x_3)\Theta^{(0)}(x_1)\Theta^{(0)}(x_2)}, \nonumber\\
        \mathcal{A}^{*} &= \ev{T_{uu}^{(0)}(x_3) O(x_1) \Theta^{(0)}(x_1)}, &
        \mathcal{B}^{*} &= \partial_{u3} \partial_{u3} \ev{O(x_3) O(x_1) \Theta^{(0)}(x_2)}, \nonumber\\
        \mathcal{A}^{**} &= \ev{ T_{uu}^{(0)}(x_3) O(x_1) O(x_2)},  &         \mathcal{B}^{**} &= \partial_{u3} \partial_{u3} 
        \ev{O(x_3) O(x_1) O_(x_2)},
\end{align}
% \begin{align}
%   &\mathcal{A}
%   % = \mathcal{A}(x_3; x_1, x_2) 
%   = \ev{T_{uu}^{(0)}(x_3) \Theta^{(0)}(x_1) \Theta^{(0)}(x_2)}, \quad
%   \mathcal{B} 
%   % = \mathcal{B}(x_3; x_1, x_2) 
%   =\partial_{u3} \partial_{u3} \ev{O(x_3)\Theta^{(0)}(x_1)\Theta^{(0)}(x_2)}, \cr
%   &\mathcal{A}^{*} = \ev{T_{uu}^{(0)}(x_3) O(x_1) \Theta^{(0)}(x_1)}, \quad
%   \mathcal{B}^{*} = \partial_{u3} \partial_{u3} \ev{O(x_3) O(x_1) \Theta^{(0)}(x_2)}, \cr
%   &\mathcal{A}^{**} = \ev{ T_{uu}^{(0)}(x_3) O(x_1) O(x_2)}, \quad
%   \mathcal{B}^{**} = \partial_{u3} \partial_{u3} 
%  \ev{O(x_3) O(x_1) O_(x_2)},
% \end{align}
and used 1$\leftrightarrow$2 symmetry. For now, let us assume that all of them are free from IR and UV divergence. Then all the correlation functions can be Taylor expanded around $k_1 = k_2 =0$. If there are more than two derivatives, they do not contribute to the sum rule because we will eventually set $k_1=k_2=0$. 

Thus, the only non-trivial contribution could come from $\mathcal{A}^*$ and $\mathcal{B}$. In $\mathcal{A}^*$, from the Lorentz invariance, $T_{uu}^{(0)}$ necessarily provides extra factor of $a (k_{1u}-k_{2u})^2 + b(k_{1u}+k_{2u})^2$ (because $g_{uu}=0$), so it is higher order in momentum. Dimensionwise, $\mathcal{B}$ is dangerous, but it is proportional to $k_{3u}^2=(k_{1u}+k_{2u})^2$.  Then, because $(\partial_{k_{1u}}-\partial_{k_{2u}})^2 (k_{1u}+k_{2u})^2 = 0$, it does not contribute, either.

The IR finiteness is guaranteed from the assumption that the theory becomes gapped. The assumption of UV finiteness is more subtle and this is one reason why we prefer this sum rule to the other ones. While we can come up with a dangerous improvement term that violates this UV finiteness (e.g. by considering a higher derivative operator), we will explicitly demonstrate the UV finiteness of the improvement terms in examples even if they spoiled Zamolodchikov's sum rule discussed in section 2.\footnote{We also note that there exist some ambiguities in treating the UV contact terms. See Appendix A for how such ambiguity may affect the three-point functions.}

Once it is a finite number, what is it? We expect it is the Virasoro central charge of the UV theory $c_{\text{UV}}.$ More precisely, since the UV central charge can be changed with the improvement term (even if we assume $\Theta_{\text{UV}}= 0$) such as in linear dilaton theory, it is the effective central charge $c_{\text{UV}}^{\mathrm{eff}}$. The effective central charge is a useful concept because it captures the actual growth of the states at high energy. In the linear dilaton theory, for instance, the effective central charge is $1$ independent of the background charge while the central charge depends on the background charge and is different from $1$.

We emphasize again that the three-point function sum rule here is not equivalent to Zamolodchikov's sum rule \eqref{Zsum}. The latter is divergent when $\Theta_{\text{UV}} \neq 0$ even if the IR theory is gapped. Also, Zamolodhikov's sum rule computes the central charge rather than the effective central charge.
The three-point function sum rule by Hartman and Mathys removes the UV divergence in a particular manner.

Let us finally ask: what happens if we drop the condition that the IR theory is gapped? Unfortunately, the IR divergence will appear and we cannot say definite things any longer about this quantity. The dimensional analysis predicts logarithmic divergence, and the sum rule leads to infinity. We can explicitly demonstrate this in the examples in section 5 by taking the $m^2 \to 0$ limit.

\section{A free scalar with arbitrary improvement terms: three-point functions}\label{sec:3ptexmple}

Our main goal is to explicitly compute the three-point function sum rule of \cite{hartman2023twodimensional} in the massive free scalar RG flow with the improvement term in the energy-momentum tensor.
The energy-momentum tensor of a free massive scalar in the Minkowski signature is defined as
\begin{align}
  T^{(\beta)}_{uu} &= -\frac{1}{\alpha'}\qty[\partial_u X \partial_u X + \beta \partial_u \partial_u f(X)] \\
  \Theta^{(\beta)} &= \frac{1}{\alpha'} [m^2 X^2 - 4\beta \partial_u \partial_v f(X)],
\end{align}

To compute the three-point function in momentum space, we will use 
the Feynman propagator of a massive scalar. In our convention, it is given by 
\begin{align}
  \mathcal{G}_{ij} = \mathcal{G}_L(x_i-x_j)= -2\pi\alpha' i \int \frac{d^2p}{(2\pi)^2} \frac{e^{ip(x_i-x_j)}}{p^2 + m^2 - i\epsilon}.
\end{align}

\subsubsection{\underline{Without the improvement term}}
We start with the momentum space three-point function without improvement studied in \cite{hartman2023twodimensional}
\begin{align}
  & \ev{\ev{T_{uu}^{(\beta)}(k_3) \Theta^{(\beta)}(k_1) \Theta^{(\beta)}(k_2)}}\nonumber \cr
  % &=-\int d^2x_1 d^2x_2 d^2x_3 e^{-ik_1x_1}e^{-ik_2x_2}e^{-ik_3x_3} 8i(2\pi)^3 m^4 \nonumber\\
  % &\quad\quad\quad\quad\quad\times\int \frac{d^2p_1 d^2p_2 d^2p_3}{(2\pi)^6}
  % \frac{p_{1u}p_{2u}e^{ix_1(p_3-p_2)}e^{ix_2(-p_3-p_1)}e^{ix_3(p_1+p_2)}}{(p_1^2 + m^2 - i\epsilon)(p_2^2 + m^2 - i\epsilon)(p_3^2 + m^2 - i\epsilon)} \nonumber\\
  &=-8i(2\pi)^3 m^4 \int d^2p \frac{(-p-k_3)_u p_u \delta(k_1+k_2+k_3)}{((p+k_3)^2 + m^2 - i\epsilon)(p^2 + m^2 - i\epsilon)((p-k_1)^2 + m^2 - i\epsilon)} \cr
    &= 2(4\pi)^2m^4\int_{0}^{1} dy_1 dy_2 dy_3 
  \frac{[(y_1 k_{3u} - y_3 k_{1u})^2 - k_{3u}(y_1 k_{3u} - y_3 k_{1u})]\delta(1- y_1 - y_2 - y_3)}{(m^2 + y_1k_3^2 + y_3 k_1^2 - (y_1 k_{3} - y_3 k_{1})^2)^2}
\end{align}
Here in the second line, we have used the Feynman parametrization and performed the $d^2p$ integral with $g_{uu} = 0$.

Expanding at low momentum and using the momentum conservation, we obtain
\begin{align}
\label{threepoint}
  \ev{\ev{T_{uu}^{(\beta)}(k_3) \Theta^{(\beta)}(k_1) \Theta^{(\beta)}(k_2)}} 
  &\sim 2(4\pi)^2 \int_{0}^{1} dy_1 \int_{0}^{1-y_1} dy_3 [(y_1 k_{3u} - y_3 k_{1u})^2 - k_{3u}(y_1 k_{3u} - y_3 k_{1u})]\nonumber\\
  &= -\frac{(4\pi)^2}{12}(k_{1u}^2 + 3k_{1u}k_{2u} + k_{2u}^2) .
\end{align}
Substituting it to \eqref{Hsum}, we reproduce their result $\Delta c = -\frac{1}{2}(\partial_{k_{1u}}-\partial_{k_{2u}})^2 (k_{1u}^2 + 3k_{1u}k_{2u} + k_{2u}^2) = 1$. 

\subsubsection{\underline{Example 1: $f=X^2$}}
Now let us consider the three-point function sum rule with non-trivial improvement terms. First, let us study the case with $f=X^2$,
\begin{align}
  &\ev{T_{uu}^{(\beta)}(x_3) \Theta^{(\beta)}(x_1) \Theta^{(\beta)}(x_2)} \nonumber\\
  &= -\frac{1}{\alpha'^3} \ev{(\partial_{u3}X_3 \partial_{u3}X_3 + \beta \partial_{u3} \partial_{u3}X_3^2)
  (m^2X_1^2 - 4\beta \partial_{u1}\partial_{v1}X_1^2)(m^2X_2^2 - 4\beta \partial_{u2}\partial_{v2}X_2^2)} \nonumber\\
  % &= \frac{8m^2}{\alpha'^3} \mathcal{G}_{12} (\partial_{u3} \mathcal{G}_{31}) (\partial_{u3} \mathcal{G}_{32}) \nonumber\\
  % &+ \frac{8\beta}{\alpha'^3} [8 \partial_{u2} \partial_{v_2}(\mathcal{G}_{12} (\partial_{u3} \mathcal{G}_{31}) (\partial_{u3} \mathcal{G}_{32}))
  % \textcolor{red}{- m^4 \partial_{u3} \partial_{u3}(\mathcal{G}_{31} \mathcal{G}_{32} \mathcal{G}_{12})}] \nonumber\\
  % &+ \frac{64\beta^2}{\alpha'^3}[m^2 \partial_{u3} \partial_{u3} \partial_{u1} \partial_{v1} (\mathcal{G}_{12} \mathcal{G}_{31} \mathcal{G}_{32}) 
  % - 2 \partial_{u1} \partial_{v1} \partial_{u2} \partial_{v2}(\mathcal{G}_{12} (\partial_{u3}\mathcal{G}_{31}) (\partial_{u3}\mathcal{G}_{32}))]\nonumber\\
  % &+ \frac{16\cdot8\beta^3}{\alpha'^3} \partial_{u3} \partial_{u3} \partial_{u1} \partial_{v1} \partial_{u2} \partial_{v2} (\mathcal{G}_{12} \mathcal{G}_{31} \mathcal{G}_{32})\nonumber\\
  &= \frac{8m^2}{\alpha'^3} A 
  + \frac{8\beta}{\alpha'^3} [8 \partial_{u2} \partial_{v_2} A - m^4 B] \nonumber\\
  &\quad\quad\quad\quad\quad
  + \frac{64\beta^2}{\alpha'^3}[m^2 \partial_{u1} \partial_{v1} B  - 2 \partial_{u1} \partial_{v1} \partial_{u2} \partial_{v2} A ]
  + \frac{16\cdot8\beta^3}{\alpha'^3} \partial_{u1} \partial_{v1} \partial_{u2} \partial_{v2} B,
\end{align}
where we put 
\begin{align}
  &A= A(x_3; x_1, x_2) = \mathcal{G}_{12} (\partial_{u3} \mathcal{G}_{31}) (\partial_{u3} \mathcal{G}_{32}), \nonumber\\
  &B = B(x_3; x_1, x_2) =\partial_{u3} \partial_{u3} (\mathcal{G}_{31} \mathcal{G}_{32} \mathcal{G}_{12}).
\end{align}
We have already calculated $A$, so we only need to calculate $B$. 
The other terms do not contribute to $\Delta c$ because they are a higher order of momentum and it becomes zero at $k_1=k_2=0$ in the momentum space.

Now we explicitly calculate $B$ in the momentum space to show that it is finite in both UV and IR, which means that the formal argument in section 5 is literally valid. The loop integral can be computed by using the Feynman parameterization: 
\begin{align}
  &-\frac{8\beta m^4}{\alpha'^3} B(k_3; k_1, k_2) \nonumber\\
  % &= -8i\beta m^4(2\pi)^3 \int\frac{d^2p_1 d^2p_2 d^2p_3}{(2\pi)^6}\int d^2x_1 d^2x_2 d^2x_3 \nonumber\\
  % &\quad\quad\quad\quad\quad\quad\quad\quad\quad\quad\times
  % \frac{-(p_1+p_2)^2 e^{i(k_1-p_2+p_3)x_1}e^{i(k_2-p_3-p_1)x_2}e^{i(k_3+ p_1+p_2)x_3}}
  % {(p_1^2 + m^2 - i\epsilon)(p_2^2 + m^2 - i\epsilon)(p_3^2 + m^2 - i\epsilon)} \nonumber\\
  % &=-8i\beta m^4(2\pi)^3 \int d^2p_1 d^2p_2 d^2p_3\frac{-(p_1+p_2)^2}{(p_1^2 + m^2 - i\epsilon)(p_2^2 + m^2 - i\epsilon)(p_3^2 + m^2 - i\epsilon)}\nonumber\\
  % &\quad\quad\quad\quad\quad\quad\quad\quad\quad\quad\times
  % \delta(k_1-p_2+p_3)\delta(k_2-p_3-p_1)\delta(k_3+ p_1+p_2) \nonumber\\
  &=-8i\beta m^4(2\pi)^3 \int d^2p \frac{-k_{3u}^2 \delta(k_1 + k_2 + k_3)}{((k_2-p)^2 + m^2 - i\epsilon)((k_1+p)^2 + m^2 - i\epsilon)(p^2 + m^2 - i\epsilon)}\nonumber\\
  &=-8\beta m^4 (2\pi)^3 \frac{1}{2\pi} 
  \int dy_1 dy_2 dy_3 \frac{k_{3u}^2 \delta(1- y_1 -y_2 -y_3)}{(-(y_1 k_1 - y_2 k_2)^2 + m^2 + y_1 k_1^2 + y_2 k_2^2)^2}\nonumber\\
  &\quad\quad\quad\quad\quad\quad\quad\quad\quad\quad\times
  (2\pi)^2 \delta(k_1 + k_2 + k_3) ,
\end{align}
where in the third line, we have introduced the Feynman parametrization and performed the integration over $p$.
%where we use $ p_2 = k_1 + p_3, \ p_1 = k_2 - p_3$ by delta function.
Expanding at low momentum, and using momentum conservation, we obtain
\begin{align}
 \frac{4}{\pi}\beta(2\pi)^3(k_1+k_2)^2\times(2\pi)^2 \delta(k_1 + k_2 + k_3).
\end{align}
As discussed in the previous section, this term does not contribute to $\Delta c$; 
\begin{align}
  \Delta c(B) = 6\beta (\partial_{k_{1u}} - \partial_{k_{2u}})^2 (k_1+k_2)^2 \Big|_{k_1 = k_2 = 0} = 0,
\end{align}
where we use $(\partial_{k_{1u}} - \partial_{k_{2u}})^2 k_3^2=0$.
%However, if we substitute $B$ to another formula of $\Delta c$, we get,
%\begin{align}
%  \Delta c(B) = 6\beta \partial_{k_{1u}}^2 (k_1+k_2)^2 = 12\beta.
%\end{align} 

\subsubsection{\underline{Example 2: $f=X^3$}}
Let us consider a little bit more non-trivial case with $f=X^3$. The three-point function can be decomposed as 
\begin{align}
  &\ev{T_{uu}^{(\beta)}(x_3) \Theta^{(\beta)}(x_1) \Theta^{(\beta)}(x_2)} \nonumber\\
  &= \frac{1}{\alpha'^3} \ev{(-\partial_{u3}X_3 \partial_{u3}X_3 - \beta \partial_{u3} \partial_{u3}X_3^3)
  (-m^2X_1^2 + 4\beta \partial_{u1}\partial_{v1}X_1^3)(-m^2X_2^2 + 4\beta \partial_{u2}\partial_{v2}X_2^3)} \nonumber\\
  &= \frac{8m^2}{\alpha'^3} \mathcal{G}_{12} (\partial_{u3} \mathcal{G}_{31}) (\partial_{u3} \mathcal{G}_{32}) \nonumber\\
  &-\frac{72\beta^2}{\alpha'^3}\qty[16
  \textcolor{red}{\partial_{u1} \partial_{v1} \partial_{u2} \partial_{v2}(\mathcal{G}_{12}^2 (\partial_{u3}\mathcal{G}_{31}) (\partial_{u3}\mathcal{G}_{32}))}
  -8m^2 \textcolor{blue}{\partial_{u3} \partial_{u3} \partial_{u1} \partial_{v1} (\mathcal{G}_{12} \mathcal{G}_{31}^2 \mathcal{G}_{32})}] .
\end{align}
We first calculate \textcolor{blue}{the third (blue) term} in the momentum space. Since it is IR and UV finite, the two-loop integration can be explicitly performed by using the Feynman parameterization:
\begin{align}
  &\mathrm{F.T.}\qty(\textcolor{blue}{\partial_{u3} \partial_{u3} \partial_{u1} \partial_{v1} (\mathcal{G}_{12} \mathcal{G}_{31}^2 \mathcal{G}_{32})})\nonumber\\
  % \xrightarrow{Fourier}&
  =&\int \frac{d^2p d^2p'}{(2\pi)^2}
  \frac{(2\pi\alpha')^4 k_{1u} k_{1v} k_{3u}^2 \delta(k_1+k_2+k_3)}
  {[(k_3+p+p')^2+m^2-i\epsilon][p^2+m^2-i\epsilon][p'^2+m^2-i\epsilon][(k_1-p-p')^2+m^2-i\epsilon]}\nonumber\\
  =&-(2\pi)^2\int_{0}^{1}d^4 y \frac{1}{(4\pi)^2}\frac{1}{1-(y_1+y_2)^2}\nonumber\\
  &\quad\quad\times\frac{(2\pi\alpha')^4 k_{1u} k_{1v} k_{3u}^2 \delta(\sum k)\delta(1-\sum y)}
  {[m^2 + k_3^2y_1 + k_1^2y_2 -(k_3y_1-k_1y_2)^2 - \frac{1-(y_1+y_2)}{1+(y_1+y_2)}(k_3y_1-y_2k_1)^2]^2}
\end{align}
Expanding it at low momentum, and using momentum conservation, we can further perform the integration over the Feynman parameters:
\begin{align}
    &-\int_{0}^{1}d^4 y \frac{1}{(4\pi)^2}\frac{1}{1-(y_1+y_2)^2}\frac{(2\pi\alpha')^4 k_{1u} k_{1v} k_{3u}^2 \delta(1-\sum y)}{m^4}\times(2\pi)^2 \delta(k_1 + k_2 + k_3)\nonumber\\
    &= -\frac{(2\pi\alpha')^4 k_{1u} k_{1v} k_{3u}^2}{m^4(4\pi)^2} \int_{0}^{1} dy_1 \int_{0}^{1-y_1} dy_2 \frac{1}{1+y_1+y_2}\times(2\pi)^2 \delta(k_1 + k_2 + k_3)\nonumber\\
    &= \frac{(2\pi\alpha')^4}{m^4(4\pi)^2}k_{1u} k_{1v} k_{3u}^2(\log2-1) \times(2\pi)^2 \delta(k_1 + k_2 + k_3) .
\end{align}
So, in total, the third term gives
\begin{align}
    \Delta c(\text{\textcolor{blue}{third term}})
    % =\frac{3\pi}{(2\pi)^2}(\partial_{k_{1u}}-\partial_{k_{2u}})^2(2\pi)^3\frac{8\beta^2}{\pi}\frac{\alpha'}{m^2}(\log2-1)k_{1u} k_{1v} k_{3u}^2
    =24\beta^2 (\log2-1) \frac{\alpha'}{m^2}(\partial_{k_{1u}}-\partial_{k_{2u}})^2 k_{1u} k_{1v} k_{3u}^2 \Big|_{k_1 = k_2 = 0} =0 .
\end{align}
Thus, \textcolor{blue}{the third term} does not contribute to $\Delta c$ because it is higher order in momentum. It is important to note that the integration over the Feynman parameters is finite, so there is neither UV nor IR divergence.

Next, we calculate \textcolor{red}{the second (red) term}, which is a little bit more complicated but we can proceed as before
\begin{align}
  &\mathrm{F.T.}\qty(\textcolor{red}{\partial_{u1} \partial_{v1} \partial_{u2} \partial_{v2}(\mathcal{G}_{12}^2 (\partial_{u3}\mathcal{G}_{31}) (\partial_{u3}\mathcal{G}_{32}))})\nonumber\\
  % =&-(2\pi\alpha')^4 \int d^2x_1 d^2x_2 d^2x_3 \int \frac{d^2p_1 d^2p_2 d^2p_3 d^2p_3'}{(2\pi)^8} \nonumber\\
  % &\quad\quad\quad\quad\quad\times \frac{k_{1u}k_{1v}k_{2u}k_{2v}p_{1u}p_{2u}e^{i(k_1+p_2+p_3+p_3')x_1}e^{i(k_2+p_1-p_3-p_3')x_2}e^{i(k_3-p_2-p_1)x_3}}
  % {(p_1^2 + m^2 - i\epsilon)(p_2^2 + m^2 - i\epsilon)(p_3^2 + m^2 - i\epsilon)(p_3'^2 + m^2 - i\epsilon)} \nonumber\\
  =&-(2\pi\alpha')^4 \int \frac{d^2p_3 d^2p_3'}{(2\pi)^2} \nonumber\\
  &\quad\quad\times \frac{k_{1u}k_{1v}k_{2u}k_{2v} (-k_2+p_3+p_3')_{u} (-k_1-p_3-p_3')_{2u} \delta(k_1+k_2+k_3)}
  {[(-k_2+p_3+p_3')^2 + m^2 - i\epsilon] [(k_1+p_3+p_3')^2 + m^2 - i\epsilon] [p_3^2 + m^2 - i\epsilon] [p_3'^2 + m^2 - i\epsilon]} \nonumber\\
  =&\frac{i(2\pi\alpha')^4}{2\pi}\int dy_1 dy_2 dy_3 dy_3' \int d^2p_3 \nonumber\\
  &\quad\times\frac{k_{1u}k_{1v}k_{2u}k_{2v}[-k_{1u}k_{2u} + (k_{1u}-k_{2u})(p_{3u}+ p_u^*) + (p_{3u}^2 + 2p_{3u}p_u^* + p_u^{*2})]\delta(1-\sum y)\delta{\sum k}}
  {\qty[(1-(y_1+y_2)^2)p_3^2 + 2[(y_2 k_1 - y_1 k_2)(1+y_1+y_2)]p_3 + m^2 +y_1 k_2^2 + k_1^2 y_2 -(y_2k_1 - y_1k_2)^2]^3}\nonumber\\
  =&-\frac{(2\pi\alpha')^4}{16\pi^2}\delta\qty(1-y_1-y_2-y_3-y_3')(2\pi)^2\delta\qty(k_1+k_2+k_2) \nonumber\\
  &\quad\times \int d^4y \frac{k_{1u} k_{1v} k_{2u} k_{2v}
  \qty[-k_{1u}k_{2u} + (k_{1u}-k_{2u}) (p_u^{**}+ p_u^*) + (p_{u}^{**2} + 2p_{u}^{**}p_u^* + p_u^{*2})]}
  {(1-(y_1+y_2)^2)^3\qty[-p^{**2} + (m^2 +y_1 k_2^2 + k_1^2 y_2 -(y_2k_1 - y_1k_2)^2)/(1-(y_1 + y_2)^2)]^2}
  \nonumber\\
  =&\frac{(2\pi\alpha')^4}{16\pi^2}\int d^4y \frac{k_{1u}k_{1v}k_{2u}k_{2v}
  k_{1u}k_{2u}\delta\qty(1-\sum_i y_i)(2\pi)^2\delta\qty(\sum_i k_i)}
  {(1-(y_1+y_2)^2)\qty[-\frac{1-(y_1 + y_2)}{1+(y_1 + y_2)}(y_2k_1-y_1k_2)^2 + m^2 +y_1 k_2^2 + k_1^2 y_2 -(y_2k_1 - y_1k_2)^2]^2},
\end{align}
where we use
\begin{align}
  &p^* = %-y_1k_2 + y_2k_1 + (y_1 + y_2)p_3\ \rightarrow\ 
  -y_1k_2 + y_2k_1 + (y_1 + y_2)p^{**} = \frac{y_2k_1-y_1k_2}{1+(y_1+y_2)} = -p^{**}\nonumber\\
  &p^{**} = -\frac{y_2k_1-y_1k_2}{1+(y_1+y_2)}.
\end{align}
Expanding at low momentum, and using momentum conservation, 
\begin{align}
  &\frac{(2\pi\alpha')^4}{(4\pi)^2}\int d^4y \frac{k_{1u} k_{1v} k_{2u} k_{2v} k_{1u} k_{2u}}
  {(1-(y_1+y_2)^2) m^4}\delta(1-\sum y)\times(2\pi)^2\delta\qty(k_1+k_2+k_2)\nonumber\\
  &=\frac{(2\pi\alpha')^4}{(4\pi)^2 m^4} (1-\log{2})k_{1u}k_{1v}k_{2u}k_{2v}k_{1u}k_{2u} \times (2\pi)^2\delta\qty(k_1+k_2+k_2) .
\end{align}
So, in total, the red term gives %$-\frac{72\beta^2}{\alpha'^3}\cdot16$,
\begin{align}
    \Delta c(\text{\textcolor{red}{second term}}) 
    =& - 3\cdot 144\beta^2 (1-\log{2}) \frac{\alpha'}{m^4}\left( \partial_{k_{1u}} - \partial_{k_{2u}} \right)^2 k_{1u}k_{1v}k_{2u}k_{2v}k_{1u}k_{2u} \Big|_{k_1 = k_2 = 0}\nonumber\\
    =& 0.
\end{align}
Thus, the \textcolor{red}{second term} does not contribute to $\Delta c$ because it is again higher order in momentum and it becomes zero at $k_1=k_2=0$. 

We can generalize the calculation with any (finite) polynomial function $f(X)$. The examples here show that the three-point function sum rule by Hartman-Mathys can give a finite result as long as the IR theory is gapped even if Zamolodchikov's sum rule diverges. 

At the same, through our explicit computations, we learn a lesson on how the IR and UV divergence can affect the conclusion. Suppose we try to set $m=0$ during our computation above, then the integration over the Feymann parameters gives an IR divergence, resulting in $0 \times \infty$. The formal argument that it is zero because we have many derivatives does not go through. While we have not seen it in our examples, we could have also encountered the UV divergence if the loop integration had worse UV behavior, spoiling the finiteness.

\section{Discussions}\label{sec:discussion}
In the classic paper \cite{Polchinski:1987dy}, when he proved that scale invariance implies conformal invariance, Polchinski assumed that the theory under consideration is compact (i.e. having a discrete dilatation spectrum) to avoid the issues with the improvement we have discussed in this paper. See also \cite{Nakayama:2020bsx,Papadopoulos:2024uvi,Papadopoulos:2024tgs} for more recent discussions. What we can say is that it is always safe to assume compactness, but it is not true that non-compact field theories are not important in physics. The Liovuille theory \cite{Nakayama:2004vk} and $\mathrm{SL}(2,\mathbb{R})$ WZW model \cite{Ribault:2014hia} (and its coset) are familiar examples that frequently appear in applications to quantum gravity, (A)dS/CFT correspondence,  (topological) string duality with non-compact Calabi-Yau spaces and so on. Not only in high energy physics but there is also a condensed matter application in the Anderson localization (see e.g. \cite{RevModPhys.80.1355} for a review) and the mathematical physics application in the geometric Langrands program (see e.g. \cite{Frenkel:2005pa} for a review). In all the examples, improvement plays a crucial role. 

What will be the prospect in higher dimensions e.g in four dimensions? The situation is more interesting because unlike in two dimensions, even with assuming the discrete spectrum, an improvement of the energy-momentum tensor is allowed if the theory possesses dimension two operators (see e.g. \cite{Polchinski:1987dy,Nakayama:2013is}). Recall that in two dimensions, we needed a dimension zero operator, which makes non-compactness necessary. In simple situations, the a-theorem, the four-dimensional version of the c-theorem, was formulated without specifying the improvement ambiguities because it would not affect the dilaton scattering amplitude \cite{Komargodski:2011vj}. Yet, the mixing with the dimension two scalars can be non-trivial as studied in \cite{Dymarsky:2014zja} when they discuss the relation between scale invariance and conformal invariance. We would like to add further here that in particle cosmology, the improvement of the energy-momentum tensor often plays a physically relevant role e.g. making the Higgs inflation possible \cite{Rubio:2018ogq}. Understanding and taming improvement in the energy-momentum tensor is essential in many ways.

With this respect, we find it intriguing to see if the sum rule based on the averaged null energy condition in four dimensions (e.g. discussed in \cite{Hartman:2023qdn,Hartman:2024xkw}) may again give us a good quantity to examine, as we did in two dimensions.  

Another direction to pursue is to generalize the monotonicity of the RG flow to certain good classes of non-unitary field theories. With the so-called PT-symmetry, the $c_{\text{eff}}$ theorem \cite{Castro-Alvaredo:2017udm} demonstrates the monotonicity of the effective central charge rather than the Virasoro central charge. In non-unitary field theories, the averaged null energy condition, as well as positive spectrum axiom, in the usual sense is violated. What can replace the averaged null energy condition in the $c_{\text{eff}}$ theorem? Non-unitary field theories and a lack of the null energy condition may play an important role in dS/CFT correspondence \cite{Hikida_2022, doi2024probingsitterspaceusing}, a notion of pseudo-entropy \cite{Nakata_2021}, and wormhole physics \cite{nakamura2025wormholenonlocal}. A better understanding of the monotonicity of the RG flow between non-unitary field theories is highly welcome.

We hope to report some progress in these directions in the future.

\section*{Acknowledgemen}
YN is in part supported by JSPS KAKENHI Grant Number 21K03581. 

\appendix

\section{Using equations of motion in three-point functions}
In the main text, we defined the UV regularization of the three-point functions such that the improvement terms do not contribute to the sum rules. If a different scheme is adopted, these improvement terms may affect the sum rules. In this appendix, we discuss another natural choice where we maximize the subtraction of contact terms by employing the equations of motion. This prescription is of particular interest because, as shown below, it yields the expected shift in the central charge for the linear dilaton background.

The prescription discussed here attempts to remove potential contact terms arising from the improvement terms in the energy-momentum tensor by using the EoM. This is in accordance with the original prescription in \cite{hartman2023twodimensional}.

\subsubsection{\underline{$f(X)=X$}}
We now evaluate the three-point function for the simplest improvement, $f(X) = X$, using the equation of motion $\partial \bar{\partial} X = \frac{1}{2}m^2 X$.
\begin{align}
    &\ev{T_{uu}^{(\beta)}(x_3) \Theta^{(\beta)}(x_1) \Theta^{(\beta)}(x_2)}_{\mathrm{EoM}} \nonumber\\
    &= \frac{1}{\alpha'^3} \ev{(-\partial_{u3}X_3 \partial_{u3}X_3 - \beta \partial_{u3} \partial_{u3}X_3)
    (-m^2X_1^2 + 4\beta \partial_{u1}\partial_{v1}X_1)(-m^2X_2^2 + 4\beta \partial_{u2}\partial_{v2}X_2)} \nonumber\\
    &= -\frac{1}{\alpha'^3} \qty[m^4\ev{\partial_{u3}X_3 \partial_{u3}X_3 X_1^2 X_2^2}+\beta^2m^4\qty(\ev{\partial_{u3}X_3 \partial_{u3}X_3 X_1 X_2} -2\ev{\partial_{u3} \partial_{u3}X_3 X_1^2 X_2})] \nonumber\\
    &= -\frac{m^4}{\alpha'^3} \qty[8 (\partial_{u3} \mathcal{G}_{31}) (\partial_{u3} \mathcal{G}_{32}) \mathcal{G}_{12} + 2 \beta^2 (\partial_{u3} \mathcal{G}_{31}) (\partial_{u3} \mathcal{G}_{32}) - 4 \beta^2 (\partial_{u3} \partial_{u3} \mathcal{G}_{31}) \mathcal{G}_{12}]
\end{align}
Following the prescription discussed in the main text without further modification, we find that the first term leads to $\Delta c = 1$. The $\beta^2$ correction from the second and third terms then yields:
\begin{align}
\label{ex}
    \ev{\ev{T_{uu}^{(\beta)}(k_3) \Theta^{(\beta)}(k_1) \Theta^{(\beta)}(k_2)}}_{\beta^2} 
    &= -\frac{2m^4 \beta^2}{\alpha'^3} (2\pi)^2 \alpha'^2 \qty[\frac{k_{1u}}{k_1^2+m^2}\frac{k_{2u}}{k_2^2+m^2} +2 \frac{k_3^2}{k_3^2 + m^2}\frac{1}{k_2^2+m^2}] \nonumber\\
    &\sim -\frac{2\beta^2}{\alpha'}(2\pi)^2 (k_{1u}k_{2u}+k_{3u}^2),
\end{align}
leading to
\begin{align}
    \Delta c = 1 + 6 \frac{\beta^2}{\alpha'}.
\end{align}
This can be compared with the central charge of the linear dilaton theory (see e.g. (2.5.2) in \cite{Polchinski:1998rq}):
\begin{align}
    c= 1+ 6\alpha'V_\mu V^\mu
\end{align}
where $V_X = \frac{\beta}{\alpha'}$.  Note that had we not used the equations of motion, we would have obtained $\Delta c = 1$, which should be interpreted as the change in the \textit{effective} central charge, as discussed in the main text. We also remark that if this prescription with the equations of motion is applied to non-linear $f(X)$ (such as $X^2$ or $X^3$), UV divergences persist, in contrast to the prescription used in the main text.

The appearance of the central charge, rather than the effective central charge, is appealing here; however, establishing this in the general case requires more careful analysis. The crucial observation is that the $\beta^2$ term in the three-point function considered in this appendix is exactly the same as the one studied in \cite{Nakamura:2025hyw} in the context of proving the $k$-theorem. There, we found that this three-point function contains partial contact terms that contribute to the quantity of interest, namely, $\Delta k$ in that context, and $\Delta c$ here. The derivation in \cite{hartman2023twodimensional} did not take these partial contact terms into account, based on the assumption that there are no other operators with $\Delta = J=2$. Since such an operator does exist in the case under discussion (i.e., $\partial^2 X$), there must be further UV contributions beyond those studied in \cite{hartman2023twodimensional} to cancel the partial contact terms. It remains an important task to establish the complete story when the operator with $\Delta = J=2$ is not unique, allowing for the improvement of the energy-momentum tensor.

With respect to the contact terms, we would like to mention a related subtlety in the UV contributions as well. Let us go back to the derivation of the three-point sum rule from the contact terms in the energy-momentum tensors. According to equations (A.22) and (A.24) in the Appendix of \cite{hartman2023twodimensional}, they are  given by
\begin{align}
    \ev{\ev{\mathcal{T} \qty[\Theta(k_1) \Theta(k_2)]}} &= -i \frac{(2\pi)^2}{12\pi} c k_1^2 \\
    \ev{\ev{\mathcal{T} \qty[\Theta(k_1) \Theta(k_2) T_{uu}(k_3)]}} &= \frac{(2\pi)^3}{6\pi} c k_{1u} k_{2u},  \label{HMEMtensor}
\end{align}
in our notation, where $T_{\mu\nu} = -2\pi T_{\mu\nu}^{\text{HM}}$\footnote{Note $X = \sqrt{2\pi\alpha'}\phi$.}.

On the other hand, if we directly calculate the $\beta^2$ contributions from simply substituting 
\begin{align}
  T^{}_{uu} &= -\frac{1}{\alpha'}\qty[\partial_u X \partial_u X + \beta \partial_u \partial_u X] \\
  \Theta^{} &= \frac{1}{\alpha'} [- 4\beta \partial_u \partial_v X],
\end{align}
and performing Wick contractions, we obtain
\begin{align}
    \ev{\ev{\mathcal{T} \qty[\Theta(k_1) \Theta(k_2)]}}_{\beta^2} &= -i\frac{(2\pi)\beta^2}{\alpha'} k_1^2 \\
    \ev{\ev{\mathcal{T} \qty[\Theta(k_1) \Theta(k_2) T_{uu}(k_3)]}}_{\beta^2} &= -\frac{2(2\pi)^2\beta^2}{\alpha'} k_{1u}k_{2u},
\end{align}
which, however, is inconsistent with \eqref{HMEMtensor}. The sign in the three-point function appears incorrect.

This suggests there should exist further contributions to the contact terms in the three-point functions beyond what we directly observe from the flat-space expression of the energy-momentum tensor. Optimistically, such contributions should also resolve the puzzle we encountered regarding the partial contact terms above. To verify this, bona fide computations of correlation functions in curved space are necessary, which we leave for future work.

%%%%%%%%%%%%%%%%%%%%%%%%%%%%%%%%%%%%%%%%%%%%%%%%%%%%%%%%
%%%%%%%%%%%%%%%%%%%%%%%%%%%%%%%%%%%%%%%%%%%%%%%%%%%%%%%%
% bibliography via BibTeX
\bibliographystyle{JHEP}
\bibliography{c_theorem}

%%%%%%%%%%%%%%%%%%%%%%%%%%%%%%%%%%%%%%%%%%%%%%%%%%%%%%%%
%%%%%%%%%%%%%%%%%%%%%%%%%%%%%%%%%%%%%%%%%%%%%%%%%%%%%%%%

\end{document}